\documentclass[prl,floatfix,showpacs,amsmath,twocolumn]{revtex4}
\usepackage{amssymb}
\usepackage{graphics}

\begin{document}

\title{Matrix Product Density Operators: Simulation of
finite-T and dissipative systems.}

\author{F. \surname{Verstraete}}
\affiliation{Max-Planck-Institut f\"ur Quantenoptik,
Hans-Kopfermann-Str. 1,
  Garching, D-85748, Germany.}
\author{J. J. \surname{Garc\'{\i}a-Ripoll}}
\affiliation{Max-Planck-Institut f\"ur Quantenoptik,
Hans-Kopfermann-Str. 1,
  Garching, D-85748, Germany.}
\author{J. I. \surname{Cirac}}
\affiliation{Max-Planck-Institut f\"ur Quantenoptik,
Hans-Kopfermann-Str. 1,
  Garching, D-85748, Germany.}

\pacs{PACS}
\date{\today}

\begin{abstract}
We show how to simulate numerically both the evolution of 1D
quantum systems under dissipation as well as in thermal
equilibrium. The method applies to both finite and inhomogeneous
systems and it is based on two ideas: (a) a representation for
density operators which extends that of matrix product states to
mixed states;  (b) an algorithm to approximate the evolution (in
real or imaginary time) of such states which is variational (and
thus optimal) in nature.
\end{abstract}

\maketitle

The physical understanding of quantum many--body systems is
hindered by the fact that the number of parameters describing the
physical states grows exponentially with the number of particles.
Thus, even for a relatively small number of particles, most of the
problems become intractable. During the last decade, however,
several numerical methods have been put forward which allow us to
describe certain many--body systems. One such method is the
so--called density matrix renormalization group (DMRG), which is
very well suited to describe 1D systems on a lattice with
short--range interactions and at zero temperature
\cite{WhitePRL,WhitePRB,DMRGBook}.  In fact, DMRG has had a very
strong impact in the field of condensed matter physics, allowing
us to describe such systems with an unprecedent degree of accuracy
and to extract their physical behavior.

Several years after its discovery, DMRG was extended to finite
temperature 1D systems \cite{finitTdmrg}. The method applies to
translationally invariant infinite systems, since in that case the
evaluation of the partition function can be recasted in terms of
finding the maximum eigenvalue of a finite matrix, which in turn
can be found using a variation of the original DMRG method to
classical 2-dimensional spin systems \cite{Nishino}. The main
restriction of the method is that it does not work well for low
temperatures and that it cannot be applied in situations in which the
number of particles is finite and/or not  homogeneous, as it is
e.g. the case of atoms in optical lattices \cite{PhysicsToday}.
Recently, time--dependent versions of the DMRG method have been
put forward \cite{Cazalilla,VidalTime,Korina,WhiteTime}.

The success of the DMRG method and its extensions is based on the
fact that the many--body states that appear in some 1D problems
can be very well described in terms of the so--called matrix
product states (MPS) \cite{RomerPRL,Fannes,DMRGperiodic}, i.e.,
\begin{equation}
  \label{MPS}
 |\psi_{\mathrm{mps}}\rangle = \sum_{s_1,\ldots,s_N=1}^d {\rm Tr}
 (A^{s_1}_1 \ldots A^{s_N}_N) |s_1,\ldots,s_N\rangle.
\end{equation}
Here, the $A$'s are matrices whose dimension is bounded by some
fixed number $D$ and $d$ is the dimension of the Hilbert space
corresponding to the physical systems. In fact, DMRG can be
viewed, both for finite and infinite dimensions, as an iterative
method that for a fixed $D$ determines the matrices whose state
$|\psi_{\mathrm{mps}}\rangle$ minimizes the energy in a
variational sense \cite{RomerPRL,DMRGperiodic}. On the other hand,
the method introduced by G. Vidal in the context of simulating the
dynamics of weakly entangled quantum systems \cite{VidalMPS} and
later developed in \cite{VidalTime,WhiteTime,Korina} prescribes a
particular way to update the matrices $A$ as a function of time.
This method can be extended  \cite{VidalMPS} to determine the
evolution of mixed states by considering them as vectors in the
$d^{2N}$--dimensional space of linear operators and thus doubling
the number of matrices $A$.

In this paper, we introduce the notion of matrix product density
operators (MPDO) which extend the MPS from pure to mixed states.  We
also present {\em the optimal} way in which the time evolution of pure
and mixed states can be approximated within these two classes of states
(MPS and MPDO). The corresponding evolution in imaginary time leads to
a very versatile finite-T DMRG algorithm, not restricted to large
temperatures or to homogeneous systems. We conclude by showing how
dissipative systems governed by master equations can be efficiently
simulated.

In \cite{DMRGperiodic,FV}, a picture was introduced to analyze MPS
and DMRG from a quantum information perspective (see Fig.\ 2 in
\cite{DMRGperiodic}). The MPS (\ref{MPS}) can be depicted as being
built up by a collection of virtual D-level systems paired in
maximally entangled states. The MPS is obtained by applying linear
maps which transform the $D^2$--dimensional Hilbert space of pairs
of virtual systems into the local Hilbert spaces associated to the
physical systems. Thus, the MPS is completely specified by these
maps which in turn can be reexpressed in terms of the matrices $A$
of Eq. (\ref{MPS}). A central result of this paper is that this
picture can be extended to describe mixed quantum states. The idea
is as follows: instead of applying a linear map to the Hilbert
space associated to pairs of virtual systems, we apply a general
completely positive map to the corresponding operator space, which
is the most general map allowed by quantum mechanics
\cite{Nielsen}. We hence define the class of MPDO as the states
which can be obtained by this procedure. More specifically, a MPDO
$\rho$ of $N$ $d$--level particles with $D_1,D_2,\cdots
D_N$-dimensional bonds is defined as
\begin{eqnarray}
  \label{MPDO}
  \rho = \sum_{s_1,s_1',\ldots,s_N,s_N'=1}^d {\rm Tr} (M^{s_1,s_1'}_1 \ldots
  M^{s_N,s_N'}_N) \nonumber\\
  \times |s_1,\ldots,s_N\rangle\langle s_1',\ldots,s_N'|,
\end{eqnarray}
where $M^{s_k,s_k'}_k$ are $D_k^2\times D_{k+1}^2$ matrices that
can be decomposed as
\begin{equation}
  \label{Mpurified}
  M_k^{s,s'}=\sum_{a=1}^{d_k} A_k^{s,a} \otimes (A_k^{s',a})^\ast.
\end{equation}
Here $d_k$ is at most $d D_kD_{k+1}$, and the matrices $A_k^{s,k}$
are of size $D_k\times D_{k+1}$. Condition (\ref{Mpurified}) is a
semidefinite constraint and is sufficient to ensure that the
associated map is completely positive. Relaxing this condition
gives rise to operators which are not necessarily positive
semidefinite (i.e., the picture introduced above also allows one
to represent general operators). Note also that for sufficiently
large $D$, Eq.~(\ref{MPDO}) includes {\em any} density operator
acting on the Hilbert space. For example, the maximally mixed
state corresponds to the choice $M^{s,s'}_k
=\delta_{s,s'}\mathbf{1}$, and a pure MPS is recovered when
$M^{s,s'}_k=A_k^{s} \otimes (A_k^{s'})^\ast$.

It is possible to express any MPDO in terms of a (pure) MPS by
defining the latter over a larger Hilbert space, i.e. by using the
concept of purification \cite{Nielsen}. To each physical system we
associate an auxiliary system (or ancilla) with a Hilbert space of
dimension $d_k$, and after choosing an orthonormal basis
$|s_k,a_k\rangle$ for these particle-ancilla pairs, we write the
corresponding MPS state as
\begin{equation}
 \label{MPPurification}
 |\Psi\rangle = \sum_{s_1,\ldots,s_N}\sum_{a_1,\ldots,a_N}
 {\rm Tr} \left(\prod_{k=1}^N A^{s_k,a_k}_k\right)
 |s_1a_1,\ldots,s_N a_N\rangle.
\end{equation}
The MPDO $\rho$ is obtained after tracing over the ancillas, i.e.
$\rho = {\rm Tr}_a (|\Psi\rangle\langle\Psi|)$. It is remarkable that
for many interesting states the ancilla's dimensions can be chosen to
be small ($d_k\simeq d$) and thus the purification
(\ref{MPPurification}) yields a very efficient representation of the
MPDO. Note also that the ancilla matrices $A_k$ in
(\ref{MPPurification}) can be easily recovered from the matrices $M_k$
by means of an eigenvalue decomposition. On the other hand, given
the matrices $M_k$, one can efficiently determine expectation values
as follows:
\begin{equation}
  \langle O_1\ldots O_N\rangle_\rho = {\rm Tr}
  (E_{1,O_1}\dots E_{N,O_N})
\end{equation}
where $E_{k,O}=\sum_{s,s'} \langle s'|O|s\rangle M_k^{s,s'}$.

In order to measure the error that we make when we approximate a
density operator $\rho_0$ by some other one $\rho$, one can use
different quantities. For instance \cite{Nielsen},
$E_F(\rho,\rho_0)=1-F(\rho,\rho_0)$, where $F$ extends the notion
of fidelity from pure to mixed states, it is defined as the
maximal overlap between all possible purifications of $\rho$ and
$\rho_0$, and it is given by $F(\rho,\rho_0)= {\rm Tr}
(\sqrt{\sqrt{\rho}\rho_0\sqrt{\rho}})$. Alternatively, we may
simply take $E_{\rm HS}(\rho,\rho_0)={\rm Tr}[(\rho-\rho_0)^2]$
which is related to the Hilbert Schmidt scalar product in the
space of operators.

If we want to determine a Hamiltonian time evolution of a mixed
state in real or imaginary time, we may instead simulate the
evolution of the purification. Starting out from the MPS
purification $\left|\Psi(0)\right\rangle$ (\ref{MPPurification}),
we take a small time step and compute the next purification
$\left|\Psi(\Delta t)\right\rangle$ exactly. Unless the evolution
is local, the dimension of the matrices $A$ will increase, and we
will have to find a MPS that approximates the exact one, as in the
method introduced in \cite{VidalTime}. However, we do this here in
an optimal way, i.e. we compute the MPS which, for given
dimensions $\{D_k\}$, maximizes the overlap with the exact state.
This ``truncated'' purification is taken as our next initial data,
and we repeat the process. At any time, the purification can be
used to reconstruct the density operator and to compute
expectation values of observables.

\begin{figure}[t]
  \centering
  \resizebox{0.6\linewidth}{!}{\includegraphics{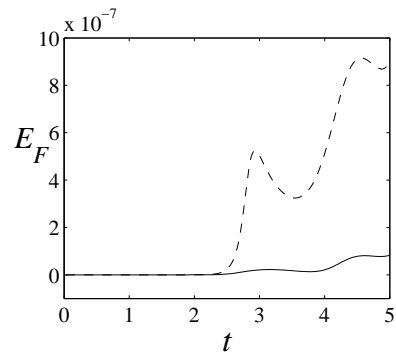}}
  \caption{
    Evolution of the ground (pure) state of an Ising model after
    switching on a small transverse magnetic field. We plot the
    infidelity, $E=1-|\langle\psi(t)|\psi_{mps}(t)\rangle|$, between
    the exact solution, $\psi(t)$, and the MPS computed using our
    iterative method (solid) and that of Ref. \cite{VidalTime} (dashed),
    using always matrix product states of dimension $D_k=4$.}
  \label{fig-error}
\end{figure}

Our method is based on an iteration which resembles the sweeps
used in standard DMRG. Let us assume that we have
$\left|\phi\right\rangle$, a MPS (\ref{MPS}) and we want to find
the closest one, $\left|\phi'\right\rangle$ in which the $A$'s are
replaced by $\tilde A$'s and the corresponding dimensions are
$\tilde D_k\le D_k$. Starting with a guess of the $\tilde A$'s
(for example, the one of the previous step), on each step of the
iteration we choose a site, $k$, and find the optimal $\tilde A_k$
such that the overlap between $\left|\phi\right\rangle$ and the
MPS is maximal. The optimal choice is given by the solution of the
system of equations
\begin{equation}
  \label{newAk}
  C_{\alpha,\beta}^{\alpha',\beta'} \tilde
  (\tilde A_k^s)_{\alpha',\beta'}
  = H^s_{\alpha,\beta}.
\end{equation}
The tensors $C$ and $D$ are defined as
\begin{eqnarray*}
  C_{\alpha,\beta}^{\alpha',\beta'} &=&
  \mathrm{Tr}(\Delta_{\alpha',\beta'}\otimes \Delta_{\alpha,\beta}
  E_{k+1}\ldots E_N E_1\ldots E_{k-1}),\\
  H^s_{\alpha,\beta} &=& \mathrm{Tr}
  (A_k^s \otimes \Delta_{\alpha',\beta'} G_{k+1}\ldots G_NG_1\ldots G_{k-1}),
\end{eqnarray*}
where $\Delta_{\alpha,\beta}$ is a matrix  with a single nonzero
element at row $\alpha$ and column $\beta$, and $E_k = \sum_s
\tilde A^s_k \otimes (\tilde A^s_k)^*$, and $G_k =\sum_s A^s_k
\otimes (\tilde A^s_k)^*$. Sweeping from $k=1$ to $N$ and back
several times we reach always a fixed point with maximal overlap
$|\langle \phi|\tilde\phi\rangle|$. The method becomes most
efficient for open boundary conditions ($D_1=D_{N+1}=1$), because
we can then always impose that on the optimized site $k$
\begin{subequations}
\begin{eqnarray}
  \sum_{s} (\tilde A^s_j) (\tilde A^s_j)^\dagger &=&
  \mathbf{1},\; \forall j > k,\mbox{ and}\label{cond1}\\
  \sum_{s} (\tilde A^s_j)^\dagger (\tilde A^s_j) &=& \mathbf{1},\;
  \forall j < k.\label{cond2}
\end{eqnarray}
\end{subequations}
Then
$C_{\alpha,\beta}^{\alpha',\beta'}=\delta_{\alpha,\alpha'}\delta_{\beta,\beta'}$
and the new matrix $A$ is just equal to $H$. In order to ensure
conditions (\ref{cond1}) and (\ref{cond2}) in the following step,
one can do use a singular value decomposition of $\tilde A^s_k$ in
exactly the same way as it is done in standard DMRG.

We have illustrated the performance of our method for a simple
case in Fig. \ref{fig-error}. We chose an Ising Hamiltonian,
$H_{z} = \sum_i^{N-1} \sigma^z_i\sigma^z_{i+1}+ \sum_i^N h
\sigma^x_i$. We begin with the ground state of $h=0$, and switch
to $h=0.2$ abruptly. We have made simulations of 6 spins with MPS
of dimension $D=4$. In Fig. \ref{fig-error} we plot the error of
our method together with that of the method introduced in
\cite{VidalTime}. Since the variational procedure is optimal at
each time step, it achieves a better result. For the cases studied
below the use of our method was crucial to obtain very good
accuracy.

\begin{figure}[t]
  \centering
  \resizebox{\linewidth}{!}{\includegraphics{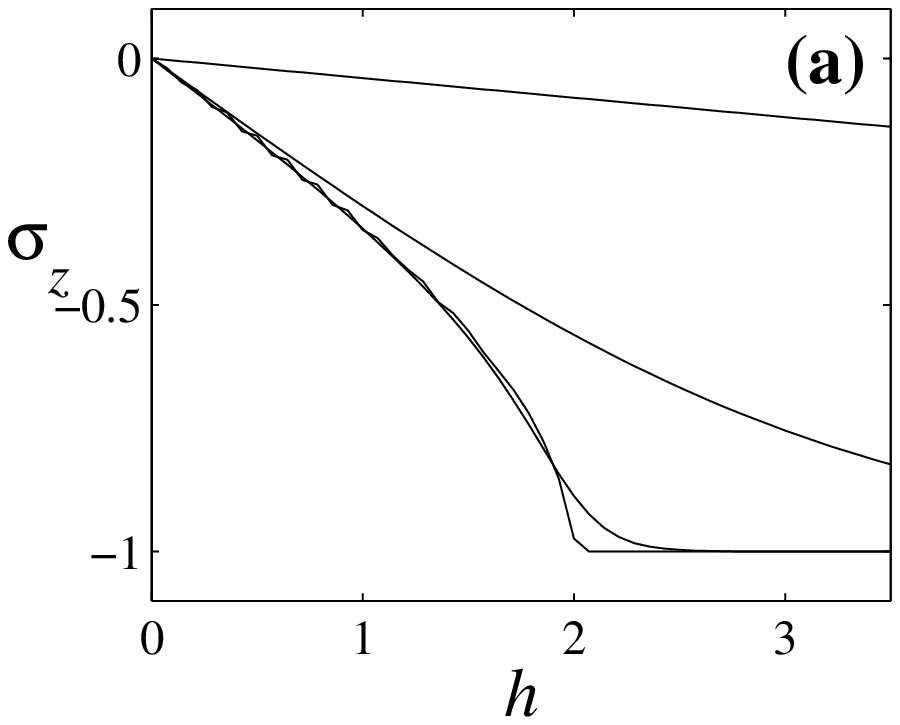}
    \includegraphics{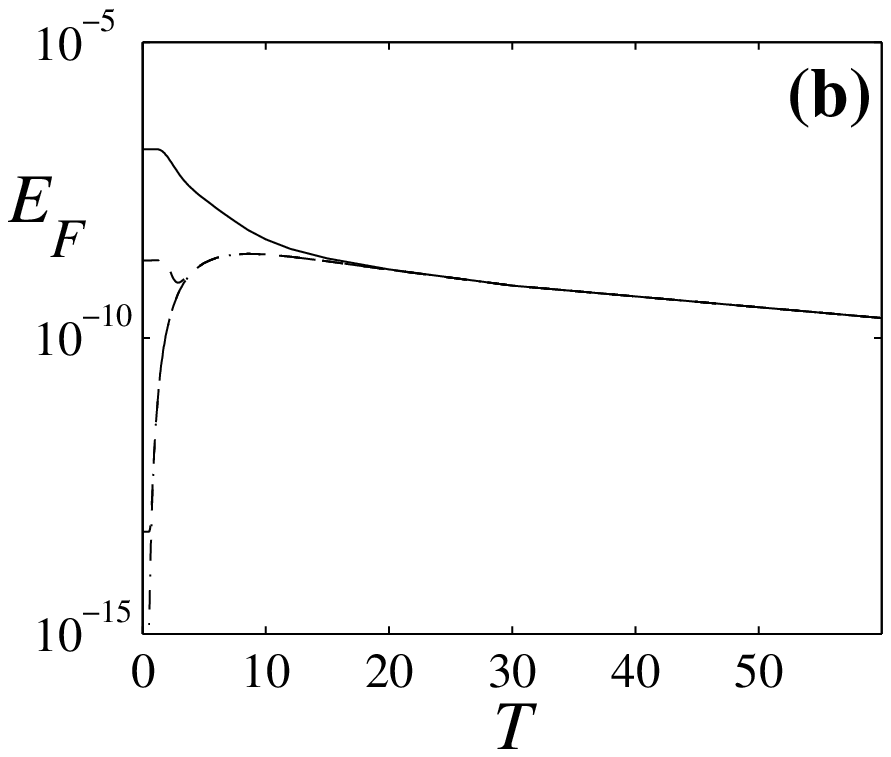}}
  \caption{
    (a) Magnetization vs.  transverse magnetic field of the thermal
    state of a XY model with $60$ spins, for temperatures
    $T=0.05,0.5,5,$ and $50$ (bottom to top).  (b) Error in the
    density matrix of the thermal state of Eq.  (\ref{xymodel})
    computed using MPDO, vs.  temperature, for a chain with $N=8$
    spins, $D=8$ (solid) and $D=14,20,24$
    (dashed). }
  \label{fig-xy}
\end{figure}

Now we apply the previous ideas to approximate the density operator
for a system at thermal equilibrium, i.e. $\rho\propto e^{-\beta
H}$ with $\beta=1/T$. For $\beta=0$ (infinite temperature), the
state $\rho$ is a maximally mixed state $\mathbf{1}$. For any
other temperature we use that \cite{VidalPC}
\begin{equation}
  e^{-\beta H} = \left(e^{-\Delta t H}\right)^M \mathbf{1}
  \left(e^{-\Delta t H}\right)^M,\;\mbox{ with }\Delta t = \tfrac{\beta}{2M}.
\end{equation}
This means that we can obtain $\rho$ by evolving according to the
Hamiltonian in imaginary time, with time step $\Delta t$. This
evolution can be performed with the optimal method presented
above, and in this case it is sufficient to fix the dimension of
the ancilla to $d$.

Using these techniques, we have computed the equilibrium states of a
spin $s=1/2$, XY model with transverse magnetic field
\begin{equation}
  \label{xymodel}
  H_{xy} = \sum_{k=1}^N \left(\sigma^x_k \sigma^x_{k+1} +\sigma^y_k\sigma^y_{k+1}
    + h \sigma^z_k\right).
\end{equation}
In Fig. \ref{fig-xy}(a) we plot the magnetization versus the magnetic
field for four different temperatures, $N=60$ spins. The accuracy of
the matrix product representation seems to increase with temperature.
Using smaller lattice sizes ($N=8$ spins) and a fixed magnetic field
($h=3.0$), we compared the exact diagonalization and the MPDO results
for different dimensions, $D$. As Fig. \ref{fig-xy}(b) shows, no large
$D$'s are needed to describe the thermal states accurately.

Let us now move on to the description of the evolution of spin
systems in the presence of decoherence. For illustrative purpose,
we consider a master equation of the form
\begin{equation}
  \label{master}
  \frac{d}{dt}\rho = i \left[H_z,\rho\right]
  + \gamma \sum_i (\sigma^x_i\rho\sigma^x_i - \rho) =: {\cal L}\rho.
\end{equation}
where $H_z$ is the Ising Hamiltonian mentioned before and $\gamma$
characterizes the interaction with the environment.

Given an initial condition $\rho(0)=\rho_0$ written in MPDO form
(\ref{MPDO}), we want to find the MPO which best approximates
$\rho(t)$. The procedure that we propose resembles the one for
Schr\"odinger equations, in that we integrate almost exactly the
master equation for a short time $\Delta t$, and then find the
optimal MPDO of smaller dimensions which is closest to
$\rho(t+\Delta t)$. Let us denote by ${\cal E}({\cal L},t) =
\exp(t {\cal L})$ the CPM so that $\rho(t) ={\cal E}({\cal
L},t)\rho(0)$. We will decompose this operator as follows
\begin{equation}
  \label{Trotter}
  {\cal E}({\cal L},\Delta t) \simeq \left[{\cal E}({\cal L}_{e},\Delta t/2)
    {\cal E}({\cal L}_{o},\Delta t)
    {\cal E}({\cal L}_{e},\Delta t/2)\right]^n,
\end{equation}
where ${\cal L} = {\cal L}_{o}+ {\cal L}_{e}$ is a splitting of the
Liouvillian into commuting terms which act on odd, $(2k+1,2k)$, and
even neighbors, $(2k,2k+1)$, respectively.

The action of the operators ${\cal E}({\cal L}_e,\Delta t/2)$ or
${\cal E}({\cal L}_o,\Delta t)$ on a state of the form
(\ref{MPDO}) can be computed exactly, and one obtains an operator
$\rho_o={\cal E}({\cal L}_o,\Delta t)\rho$ of the same form but
with the substitution $M_{2k}^{s_{2k},s_{2k}'}
M_{2k+1}^{s_{2k+1},s_{2k+1}'} \rightarrow
M^{s_{2k},s_{2k+1},s_{2k}',s_{2k+1}'}_{2k,2k+1}$. Now we find the
optimal operator $\rho_L$, which has the form of Eq.\
(\ref{MPDO}), and is built from matrices of fixed dimension. As
before, by optimal we mean that the distance to the original one
is minimized. This time we will use as a measure of the distance
$E_{\rm HS}(\rho_o,\rho_L)$ and thus we will not impose the
condition (\ref{Mpurified}) (the use of $E_{F}$ in this context is
more complicated and will be explored in a future work). This
leads to an optimization algorithm which is very much like the one
introduced before at the level of purification, but in which now
$\rho_o$ and $\rho_L$ are treated as vectors on a ${\bf d}^{N}$
dimensional space, where ${\bf d} = d\times d$. This way, it is
possible to use the same code for studying the evolution of both
pure and mixed states. We remark that, while this method cannot
guarantee the positivity of the truncated operator $\rho_L(t)$
obtained at the end of the algorithm, if the truncation error
$E_{\rm HS}(\rho_o,\rho_L)$ is kept small, it will be possible to
bound the errors made when using $\rho_L(t)$ to compute
expectation values of any physical quantity. Note also that the
matrices $M^{s_k,s_k'}_k \in \mathbb{C}^{L_k\times L_k}$ do not
need to have a size $L_k=D_k^2$ (which is the square of a number).

\begin{figure}[t]
  \centering
  \resizebox{\linewidth}{!}{\includegraphics{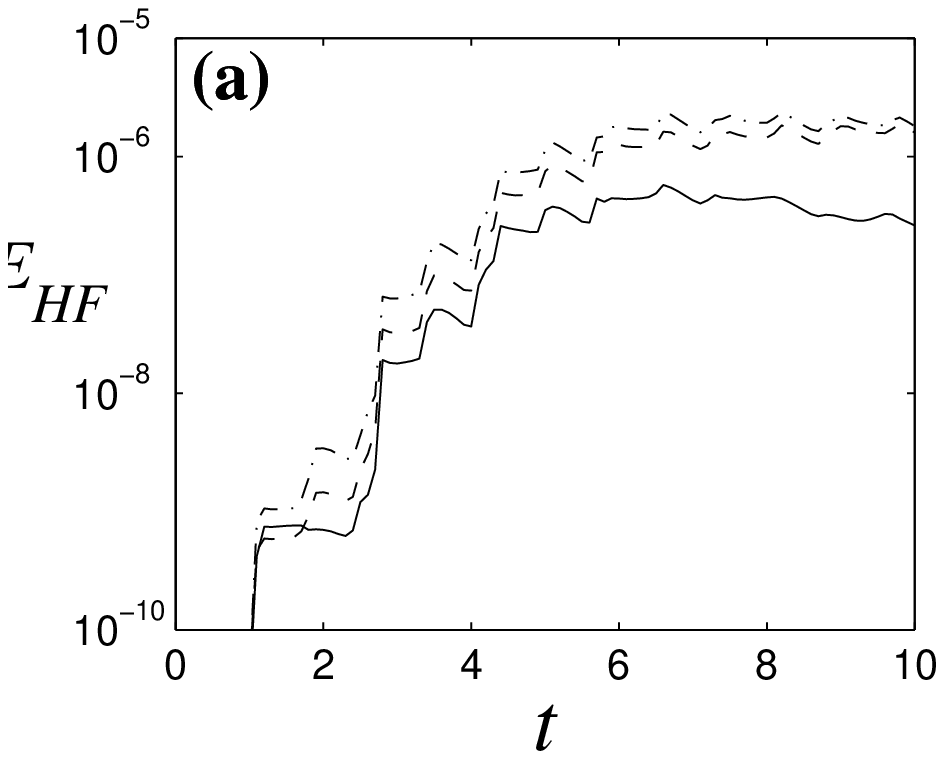}
    \includegraphics{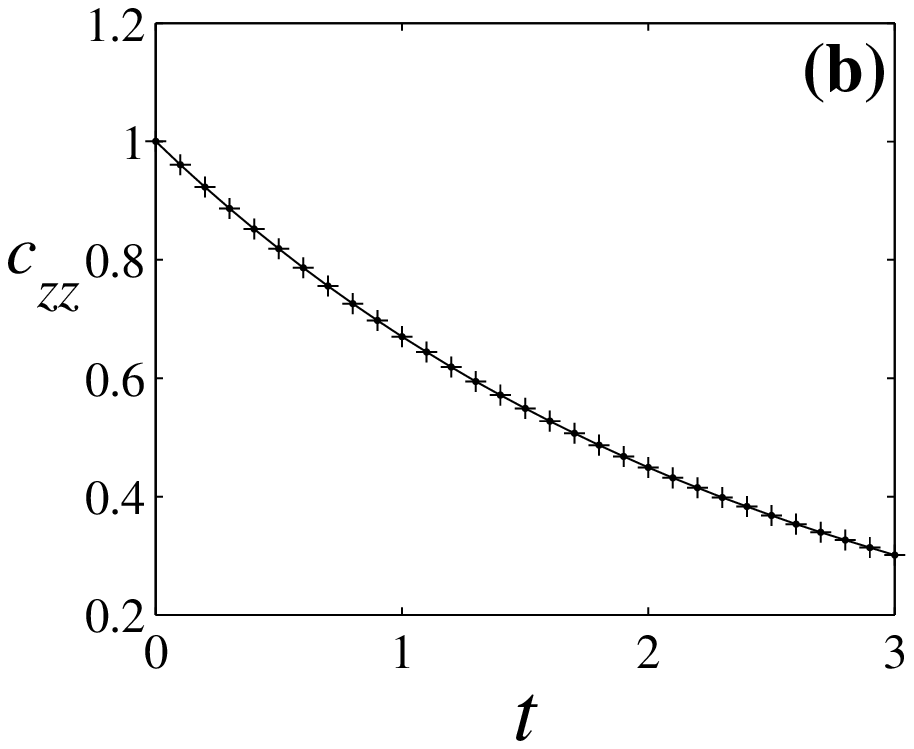}}
  \caption{(a) Errors in simulations of a cluster state with
    $N=6,7$ and $8$ spins (solid, dash and dash-dot), under Eq.
    (\ref{master}), for $\gamma = 0.1$, $h=0$, matrices of size $L=14$ and
    timestep $\Delta t = 0.01$.  (b) Decay of correlations,
    $c_{zz}=\langle \sigma^z_{i}\sigma^z_{i+1}\rangle -
    \langle\sigma^z_i\rangle \langle\sigma^z_{i+1}\rangle$, for the site
    $i=N/2$, on GHZ states with $N=10, 20$ and $30$ spins (dots, crosses and
    solid line) for $L=10$.}
  \label{fig-master}
\end{figure}

As an illustration, we have simulated the decay of correlations of
the GHZ state, $|\psi(0)\rangle =
\tfrac{1}{\sqrt{2}}(|0\rangle^{\otimes N}+|1\rangle^{\otimes N})$
and of a cluster state \cite{cluster}, under the evolution
dictated by Eq.\ (\ref{master}). In Fig. \ref{fig-master}(a) we
show the errors for simulations with up to $N=8$ spins, a size for
which we can compute the solution numerically with direct methods
(a Crank-Nicholson which already has an error of the order of
$10^{-10}$). There are two sources of errors in our method. One is
the Trotter expansion (\ref{Trotter}), but as it is of order
${\cal O}(\Delta t^2)$, it can be decreased by making shorter time
steps. The second source of error is the truncation of the MPDO to
a lower dimension. In the problems that we have simulated,
truncation error only affected the cluster state and did not grow
much with the size of the system. In Fig. \ref{fig-master} (b) we
plot some simulations made with GHZ states of up to $50$ spins,
from which an exponential decay law for the correlation functions,
$c_{zz}(t) = \langle \sigma^z_{i}\sigma^z_{i+\Delta}\rangle -
\langle\sigma^z_i\rangle \langle\sigma^z_{i+\Delta}\rangle \simeq
e^{-4\gamma t}$, can be extracted. The method is rather efficient
and can easily handle this and bigger problems in ordinary
computers.

In summary, we have introduced the concept of matrix product
density operators (MPDO) and their purification in terms of matrix
product states (MPS) in order to describe mixed states in quantum
many--body systems. We have also developed a method to determine
the MPS (MPDO) which optimally describes another one which is
composed of matrices of higher dimensions. We have used those
ideas to introduce an algorithm which allows us to study 1D
systems in thermal equilibrium. In contrast to previous methods,
it applies to {\em finite} systems, with {\em general} short range
interactions (not necessarily translationally invariant) and works
over the whole range of temperatures $T\in[0,\infty)$. In fact,
the algorithm is a very natural extension of DMRG to finite
temperatures because for $T\to 0$ one recovers the same MPS that
one would obtain with standard DMRG. We have also shown how to
simulate the evolution of a system in the presence of dissipation,
i.e. when it evolves according to a master equation which involves
short range interactions. We remark that in this last case an
alternative possibility is to combine the concepts introduced here
with the ``quantum jump'' approach \cite{noise}. The idea is to
perform several realizations of the evolution starting out with
randomly chosen pure states in such a way that their mixture
reproduces the initial MPDO (\ref{MPDO}). In each realization the
pure state is evolved according to a stochastic Schr\"odinger
equation consisting of a time evolution with an effective
Hamiltonian which is interrupted by quantum jumps \cite{noise}.
After each short time step (or quantum jump) the new pure state is
determined using the method introduce here and which optimally
approximates the new state with a MPS. At the end, expectation
values of physical observables are determined by averaging with
respect to the different realizations. Which of these two methods
is more efficient will depend on the number of realizations one
has to perform in practice in the latter in order to achieve a
prescribed accuracy.

We thank M. A. Martin-Delgado and G. Vidal for discussions. Work
supported by the DFG (SFB 631), the european project and network
Quprodis, RESQ and Conquest, and the Kompetenznetzwerk der Bayerischen
Staatsregierung Quanteninformation.

\end{document}